\documentclass{appolb}

  \usepackage{amssymb}
   \usepackage{amsmath}
    \usepackage{amsfonts}
     \usepackage{epsfig}
      \usepackage{bm} 
      \usepackage[dvipsnames,usenames]{color}

\usepackage[colorlinks,citecolor=blue,urlcolor=blue,linkcolor=blue]{hyperref}

\begin{document}

\title{L\'evy imaging of elastic hadron-hadron scattering:\\ Odderon and inner structure of the proton}

\author{T. Cs\"{o}rg\H{o}$^{1,2,3}$, R. Pasechnik$^{4}$ and A. Ster$^{1}$
\address{
{$^1$\sl 
MTA WIGNER FK, H-1525 Budapest 114, POB 49, Hungary
}\\
{$^2$\sl 
EKU KRC, H-3200 Gy\"ongy\"os, M\'atrai \'ut 36, Hungary
}\\
{$^3$\sl 
CERN, CH-1211 Geneva 23, Switzerland
}\\
{$^4$\sl
Department of Astronomy and Theoretical Physics,\\ Lund
University, SE-223 62 Lund, Sweden
}}
}

\maketitle

\begin{abstract}
A novel model-independent L\'evy imaging method is employed for reconstruction of the elastic $pp$ and $p\bar p$ scattering amplitudes at low and high energies. The four-momentum transfer $t$ dependent elastic slope $B(t)$, the nuclear phase $\phi(t)$ as well as the excitation function of the shadow profile $P(b)$  have been 
extracted from data 
at ISR, Tevatron and LHC energies. We found qualitative differences in properties of $B(t)$ and $\phi(t)$ between $pp$ and for $p\bar p$ collisions, 
that indicate an Odderon effect.  A proton substructure has also been identified and found to have two different sizes, comparable to that of a dressed quark at the ISR and of a dressed diquark at the LHC energies, respectively.
\end{abstract}


\section{Introduction}
\label{Sec:Introduction}

Recently, new data on the total, elastic and differential cross-section measurements in elastic $pp$ collisions $\sqrt{s} = 13$ TeV have become available from the TOTEM Collaboration 
at the Large Hadron Collider (LHC)~\cite{Antchev:2017dia,Antchev:2017yns} indicating a possible effect of the C-odd three-gluon state known as Odderon~\cite{Lukaszuk:1973nt}. These data have triggered an intense debate in the literature, see e.g.~refs.~\cite{Samokhin:2017kde}--\cite{Csorgo:2018uyp}. Here, we contribute to this discussion with the results of our  model-independent analysis of the most recent TOTEM data set at the highest currently available energy of $\sqrt{s} = 13$ TeV~\cite{Antchev:2018edk}. 


\section{Model-independent L\'evy imaging method}
\label{Sec:Levy}

Our imaging method is 
based on a recently found, model independent Levy expansion.
For a detailed presentation of this novel and sophisticated imaging technique, see ~\cite{Csorgo:2018uyp}, while for a brief summary of the main results, see~\cite{Csorgo:2018ruk}.
In this manuscript  we highlight only some of the most interesting results of this approach. This method is a straightforward generalisation of the well-known Laguerre, Edgeworth \cite{Csorgo:2000pf} and L\'evy~\cite{DeKock:2012gp} expansions proposed for characterisation of the nearly L\'evy stable source distributions typically used in studies of particle correlations and femtoscopy. 

We proposed to apply the L\'evy expansion (with complex coefficients) at the level of the elastic amplitude~\cite{Csorgo:2018uyp,Csorgo:2018ruk}, to have a positive definite expression for the differential cross-section of elastic scattering 
This way, we obtained the following representation of 
the elastic amplitude $T_{el}(\Delta)$
\begin{eqnarray}
\frac{d\sigma}{dt} = \frac{1}{4\pi}|T_{el}(\Delta)|^2 \,, \quad
T_{el}(\Delta) &=& i\sqrt{4\pi A w(z|\alpha)}\, 
		\left[1+\sum_{i = 1}^\infty c_i l_i (z|\alpha) \right] \, , 
\label{e:dsigmadt-Tel}
\end{eqnarray}
where $\Delta=\sqrt{|t|}$, the four-momentum transfer is denoted by $t=(p_1 - p_3)^2<0$  and $ w(z|\alpha)  =  \exp(-z^\alpha)$ stands for  the L\'evy weight known also as the stretched exponential distribution, $z=|t| R^2$ is the dimensionless positively-definite scaling variable, constructed from 
$|t|$ and a L\'evy scale parameter $R$ (represented in units of fermi in what follows), $c_j=a_j + i b_j$ are the complex coefficients of the Levy expansion, and $l_j(z|\alpha)$ stands for the orthonormal set of  L\'evy polynomials. The orthonormal L\'evy polynomials introduced above are  detailed in ref.~\cite{Csorgo:2018uyp}. This Levy expansion method leads to simple expressions for the total, elastic and differential cross-sections, as well as for the $t$ dependent nuclear slope $B(t)$ and the ratio of the real to imaginary part of the forward scattering amplitude $\rho(t)$ and the nuclear phase $\phi(t)$,
as summarized briefly in ref.~\cite{Csorgo:2018ruk} and explained in greater detail in ref.~\cite{Csorgo:2018uyp}.

In our approach, the stretched exponential $w(z|\alpha)$ characterises the degree of non-exponentiality in the elastic scattering cross-section $d\sigma/dt \propto \exp\left(-(R^2 |t|)^\alpha\right)$
at small $|t| \ll R^{-2}$ below the diffraction minimum i.e. in the diffractive cone. The non-exponentiality, even though subtle, has been confirmed by the TOTEM analysis 
of $\sqrt{s}= 8$ TeV elastic $pp$ scattering data as a more than 7$\sigma$ effect~\cite{Antchev:2015zza}. This means that in the conventional exponential parameterisation
$\frac{d\sigma}{dt} = A \exp(- B |t|)$ the diffractive slope $B$ is expected to be a non-trivial function of $|t|$ whose deviation from a constant is a relevant measure of 
non-exponentiality. 

\section{Fit results}
\label{Sec:Fits}

\begin{figure}[!htb]
\begin{center}
\begin{minipage}{0.45\textwidth}
 \centerline{\includegraphics[width=1.0\columnwidth]{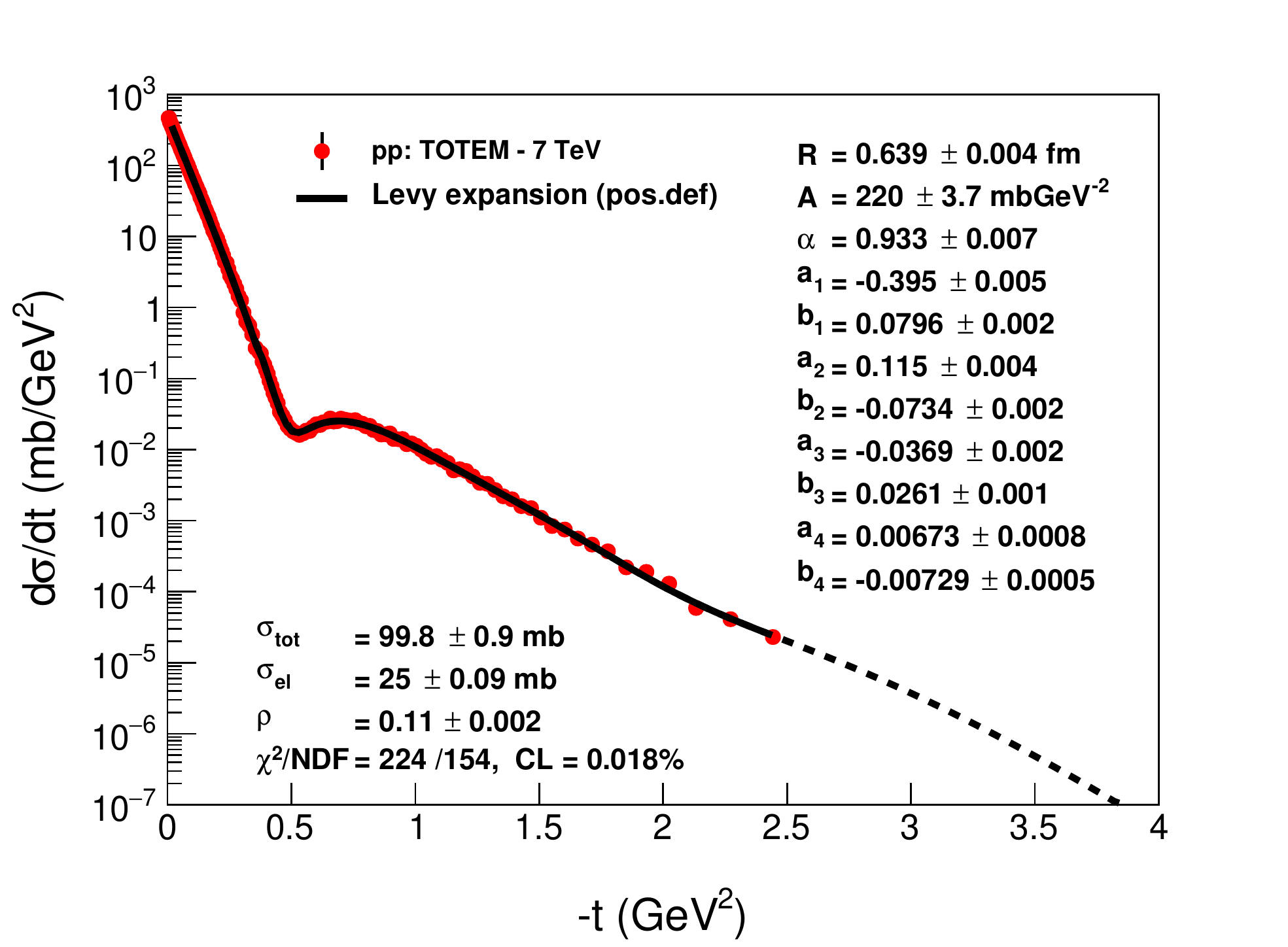}}
\end{minipage}
\begin{minipage}{0.45\textwidth}
 \centerline{\includegraphics[width=1.0\columnwidth]{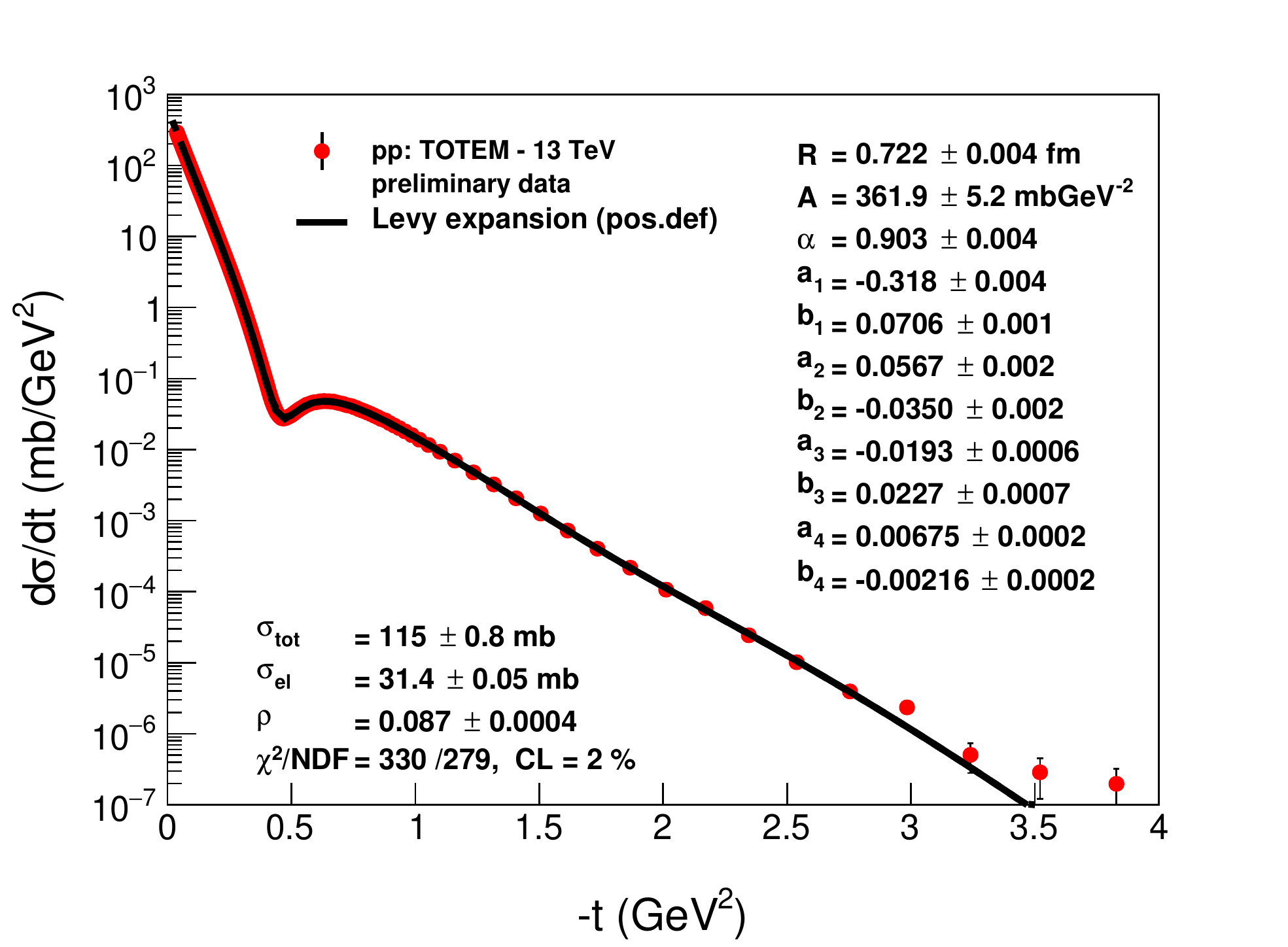}}
\end{minipage}
\end{center}
\caption{L\'evy series fits to elastic $pp$ scattering data by the TOTEM Collaboration at 
        the LHC energy of $\sqrt{s} = 7$ TeV (left panel), and  $ 13$ TeV (right panel).
}
\label{f:dsigmadt-levyfit}
\end{figure}

In Fig.~\ref{f:dsigmadt-levyfit}, we present an example of the quality fits we have performed for the differential elastic cross-section adopting the fourth-order L\'evy expansion of the elastic amplitude (\ref{e:dsigmadt-Tel}). Generically, as a necessary condition for interpretations of such fits, we consider a statistically acceptable confidence level of CL $> 0.1 \%$. As detailed in ref.~\cite{Csorgo:2018uyp}, such good quality fits were achieved for almost all of the published data sets, for both $pp$ and $p\bar p$ cases. In case of the 7 TeV $pp$ data, we obtained a marginal confidence level of CL $\approx 0.02 \%$, as indicated in
Fig.~\ref{f:dsigmadt-levyfit} (left panel). While one may try to use this fit result for physics interpretation as well, one may need to repeat the analysis of 7 TeV $pp$ data including the full
covariance matrix, and/or to study the L\'evy fits to the low-$t$ and the large-$t$ part of the 7 TeV dataset separately. 
Another possibility is to add the statistical and systematic errors in quadrature at 7 TeV, similarly to 
our analysis of the 13 TeV  data. 

\begin{figure}[!htb]
\begin{center}
\begin{minipage}{0.45\textwidth}
 \centerline{\includegraphics[width=1.0\columnwidth]{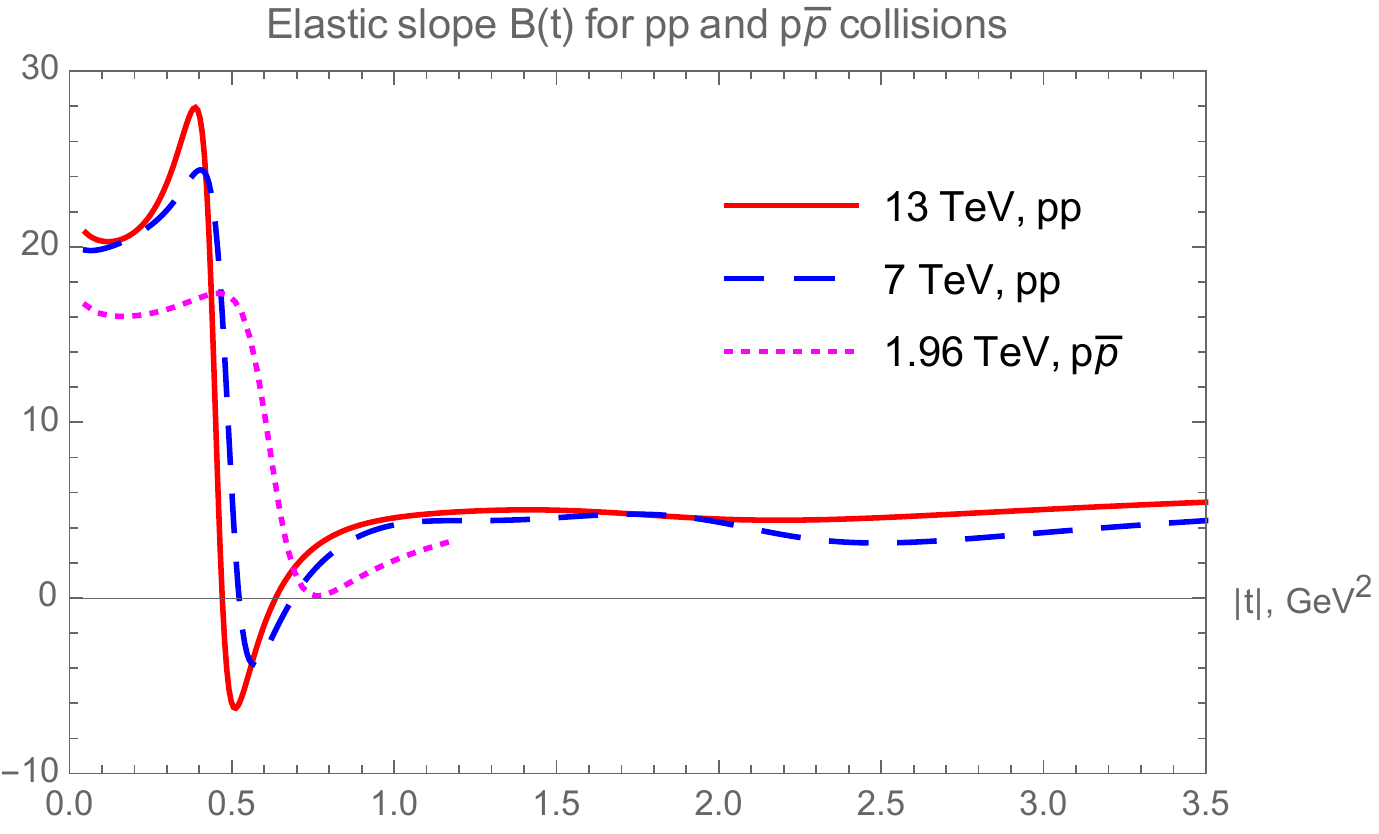}}
\end{minipage}
\begin{minipage}{0.45\textwidth}
 \centerline{\includegraphics[width=1.0\columnwidth]{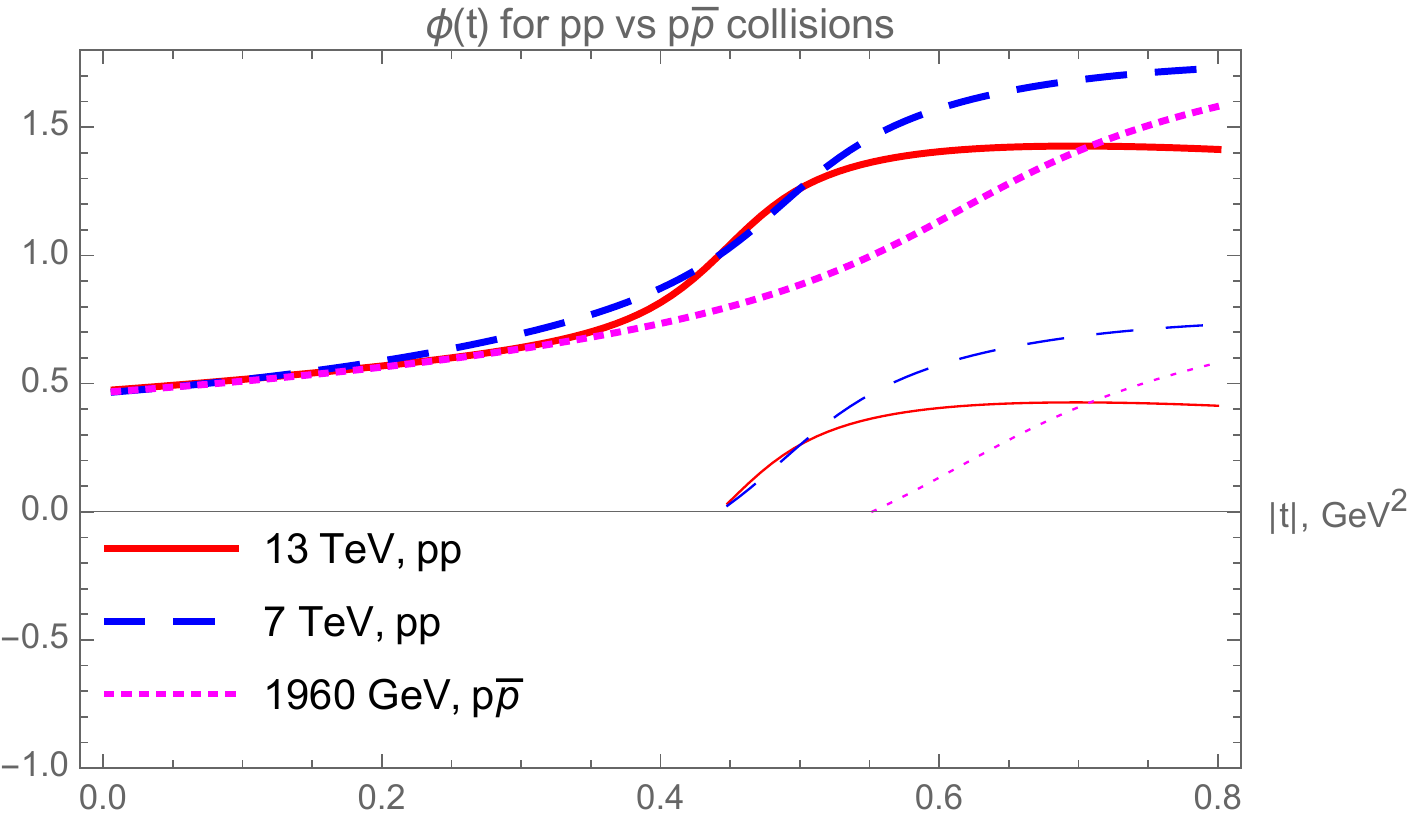}}
\end{minipage}
\end{center}
\caption{
The elastic slope $B(t)$ (left panel) and the nuclear phase $\phi(t)$ (in units of $\pi$, right panel) as functions of $t$ for three distinct energies of $pp$ and $p\bar p$ collisions.
}
\label{f:B-phi}
\end{figure}

\subsection{Characteristics of elastic scattering}
\label{Sec:Functions}

One of the most important characteristics of the elastic scattering processes
is the $t$-dependent elastic slope $B(t)$ defined as $B(t) \equiv d/dt\left(\ln d\sigma/dt \right)$.
One is interested in 
$B = B(t=0)$, which requires an extrapolation of the measured differential
cross-sections to the $t=0$ optical point. This is typically done by using an exponential
approximation.
As the L\'evy series for the elastic amplitude (\ref{e:dsigmadt-Tel}) is not an analytic function at $t = 0$ for $\alpha < 1$,  our $B(t=0)$ is a divergent quantity, if $\alpha < 1$. However, $B(t)$ exists at the finite, measurable  $t<0 $ values, and it is this $B(t)$ function that can be compared to the experimental determinations of $B$, paying also attention to the range of $t$ where the nuclear slope parameters have been experimentally evaluated. The elastic slope function $B(t)$ is shown for $pp$ ($\sqrt{s}=$ 7 and 13 TeV) and $p\bar p$ ($\sqrt{s}=$ 1.96 TeV) elastic scattering on the left panel of  Fig.~\ref{f:B-phi}.

\subsection{\it Odderon effects.}

The $pp$ scattering exhibits a well-defined dip-and-bump structure which results in crossings of the $B(t)$ function through the $B(t) = 0$ line. After the second zero-crossing, $B(t)$ in $pp$ reaches an extended and surprisingly flat plateaux in the tail regions of large $|t|$. This feature is an indication of a small substructure inside the proton, with a characteristric size and cross-section
that is detailed in ref.~\cite{Csorgo:2018ruk}.
In contrast, in $p\bar p$ collisions at Tevaton with $\sqrt{s} = 1.96$ TeV, we do not observe any dip-and-bump structure, instead, one sees a shoulder-like structure. Indeed, the corresponding $B(t)$ function does not cross zero in a statistically significant manner. Such a robust qualitative difference between the differential cross-sections of elastic  $pp$ and $p\bar p$ scattering  is a signature of the Odderon exchange, as detailed in refs.~\cite{Csorgo:2018uyp,Csorgo:2018ruk}.
The nuclear phase $\phi(t)$ is  defined as
$T_{el}(t) = |T_{el}(t)| \exp( i \phi(t) )$, corresponding to the right panel of  Fig.~\ref{f:B-phi}. Thinner lines represent the principal value of $\phi(t)$, with $0 \le \phi_{\rm PV} < \pi$.
 Fig. ~\ref{f:B-phi} shows that $\phi(t)$ reaches the value of $\pi$ at the same value of $|t|$ in $pp$ collisions both at $\sqrt{s}=$ 7 and at 13 TeV, but at a different value in $p\bar p$ collisions at $\sqrt{s}=$ 1.96 TeV, proposed as a second signature of the Odderon~\cite{Csorgo:2018uyp}.

\subsection{\it (Sub)structure(s) of the proton from elastic scattering}

\begin{figure}[htb]
\begin{center}
\begin{minipage}{0.45\textwidth}
 \centerline{\includegraphics[width=1.0\columnwidth]{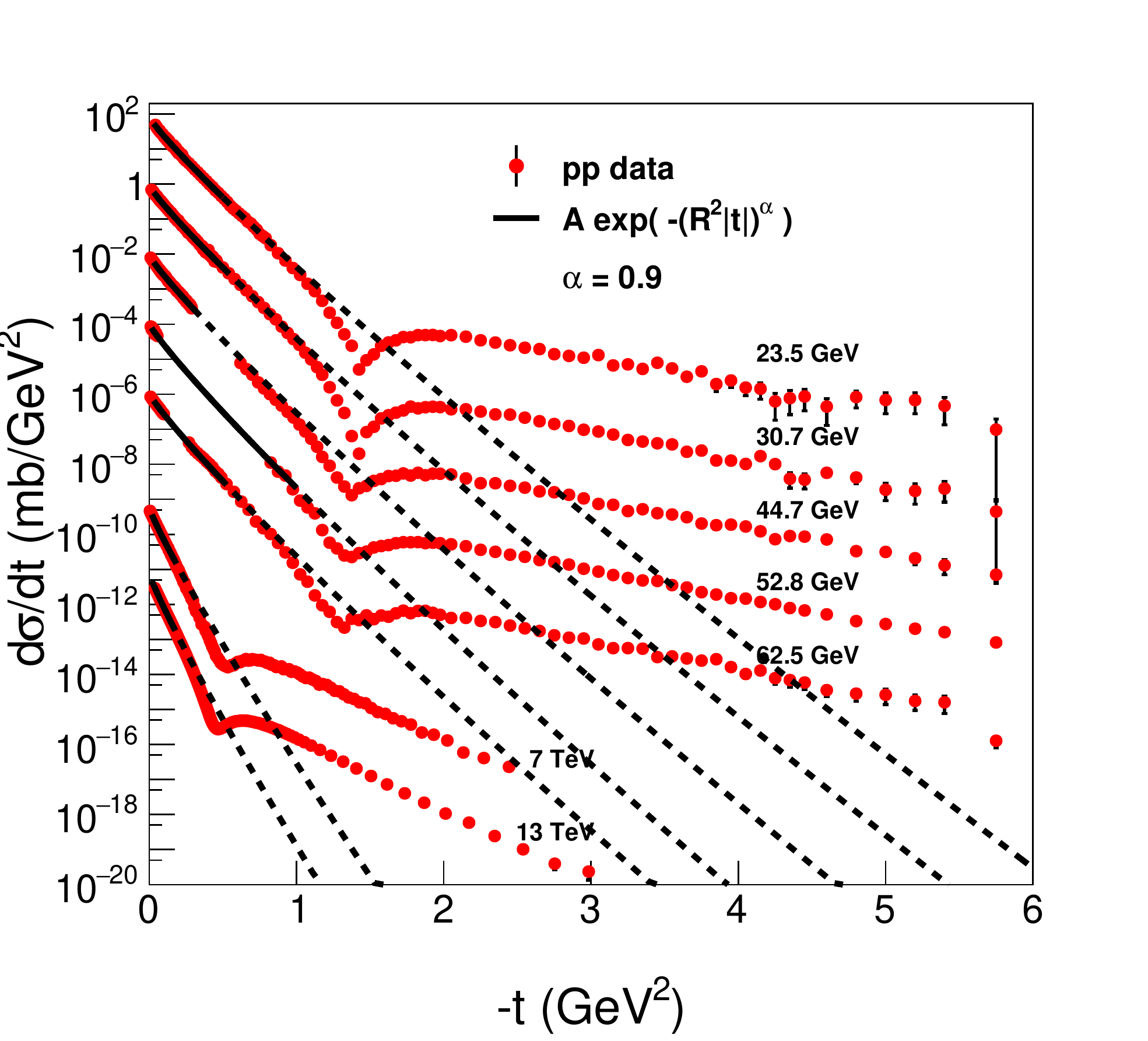}}
\end{minipage}
\begin{minipage}{0.45\textwidth}
 \centerline{\includegraphics[width=1.0\columnwidth]{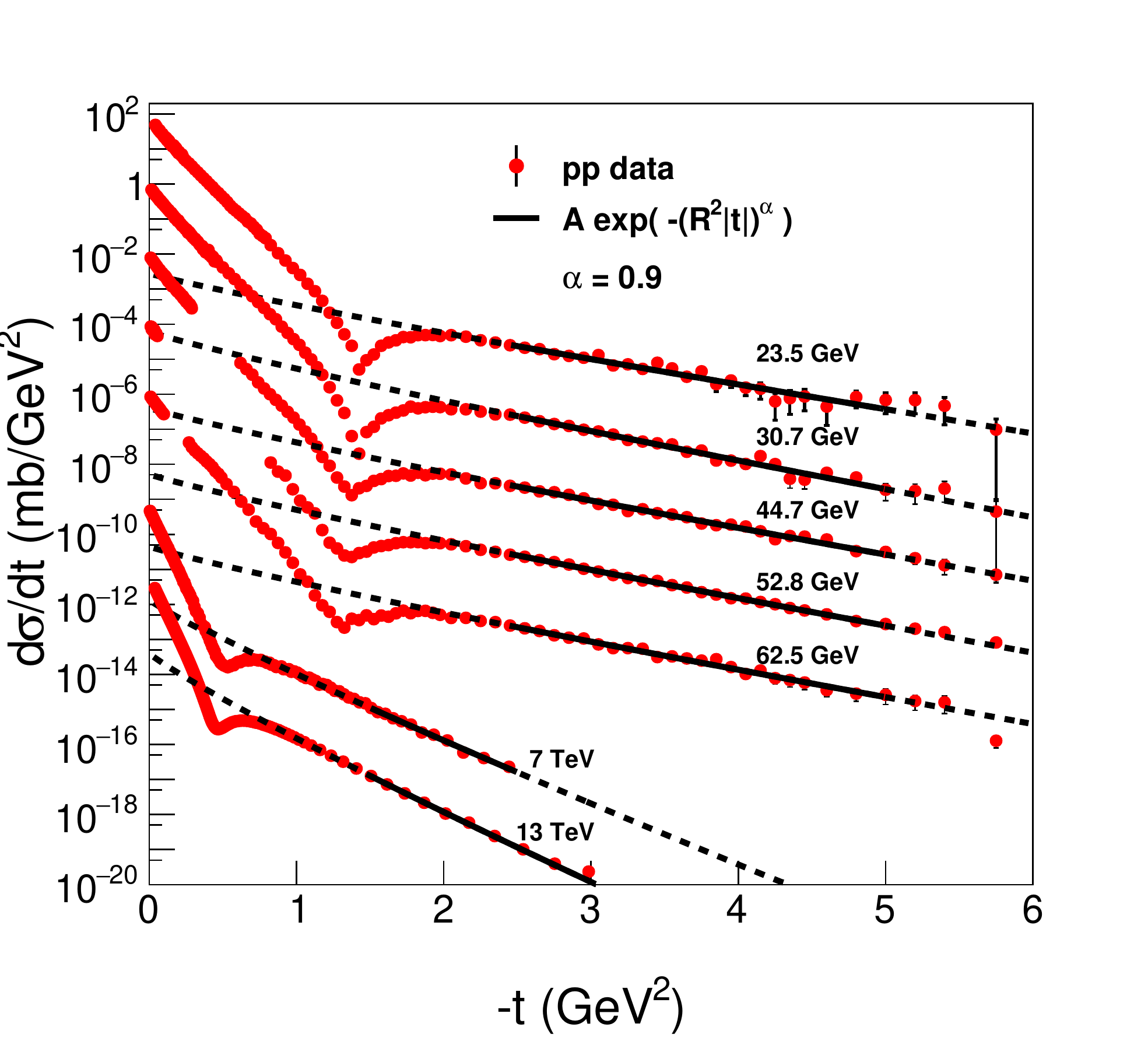}}
\end{minipage}
\end{center}
\caption{Summary plots of the zeroth-other L\'evy fits, $d\sigma/dt = A \exp(-(R^2 |t|)^{\alpha})$,
to the diffractive cone (left panel) and to the tail (right panel) regions of the elastic $pp$ scattering 
data from $\sqrt{s} = 23.4$ GeV to $13$ TeV,
with $\alpha = 0.9$ fixed.
}
\label{f:dsigmadt-cones-tails}
\end{figure}

The summary plots of the zeroth-order $d\sigma/dt = A \exp(-(R^2 |t|)^{\alpha})$ L\'evy fits  are shown on Fig.~\ref{f:dsigmadt-cones-tails}. The left and right panels indicate the leading order shape of the 
diffractive cone at low-$t$, as well as the tail or large $-t$ regions, respectively.  Both regions are well described  with an  $\alpha = 0.9$ fixed value. 
The overall size of the protons is increasing with increasing colliding energies, as evidenced by the steepening of the zeroth order L\'evy fits on the left panel of Fig.~\ref{f:dsigmadt-cones-tails}. However, the tail fits on the right panel of Fig.~\ref{f:dsigmadt-cones-tails} show parallel lines at the ISR energies, indicating a sub-structure with nearly unchanged size in the $\sqrt{s}= 23.5 - 62.5$ GeV energy range.
At the LHC energy range of $\sqrt{s} = 7 - 13$ TeV, the tail fits again show two lines that are parallel to one another, indicating a sub-structure at LHC energies, too. However, these lines are not parallel with but significantly steeper as compared to the fits to the tail of the ISR datasets. In ref.~\cite{Csorgo:2018uyp}, these results were interpreted in terms of  proton substructure(s). At the lower ISR energies, a smaller substructure with a nearly energy independent size is evidenced by the slope $B_{\rm tail}(pp|{\rm ISR}) \approx 2$ GeV$^{-2}$. At the LHC energies of $\sqrt{s} =$ 7 and 13 TeV, the slope at large $-t$ changes dramatically to $B_{\rm tail}(pp|{\rm LHC}) \approx 5 $ GeV$^2$, indicating a larger sub-structure of the protons at these energies.

\subsection{\it Proton profiles}

Let us switch to the impact parameter space as
\begin{equation}
    t_{el}(b) \, = \, \int \frac{d^2\Delta}{(2\pi)^2}\, e^{-i{\bm \Delta}{\bm b}}\,T_{el}(\Delta)\, = \, i\left[ 1 - e^{-\Omega(b)} \right]\,, \quad \Delta\equiv|{\bm \Delta}|\,, \quad b\equiv|{\bm b}|\, .\nonumber
	\label{tel-b} 
\end{equation}
The opacity $\Omega(b)$ is a complex function that defines 
the shadow profile as 
\begin{eqnarray}
	P(b) = 1-\left|e^{-\Omega(b)}\right|^2 = 
	[2-{\rm Im}\,t_{el}(b)]{\rm Im}\,t_{el}(b) - [{\rm Re}\,t_{el}(b)]^2 \,.
    \label{e:shadow}
\end{eqnarray}
\begin{figure}[!h]
 \centerline{\includegraphics[width=0.45\columnwidth]{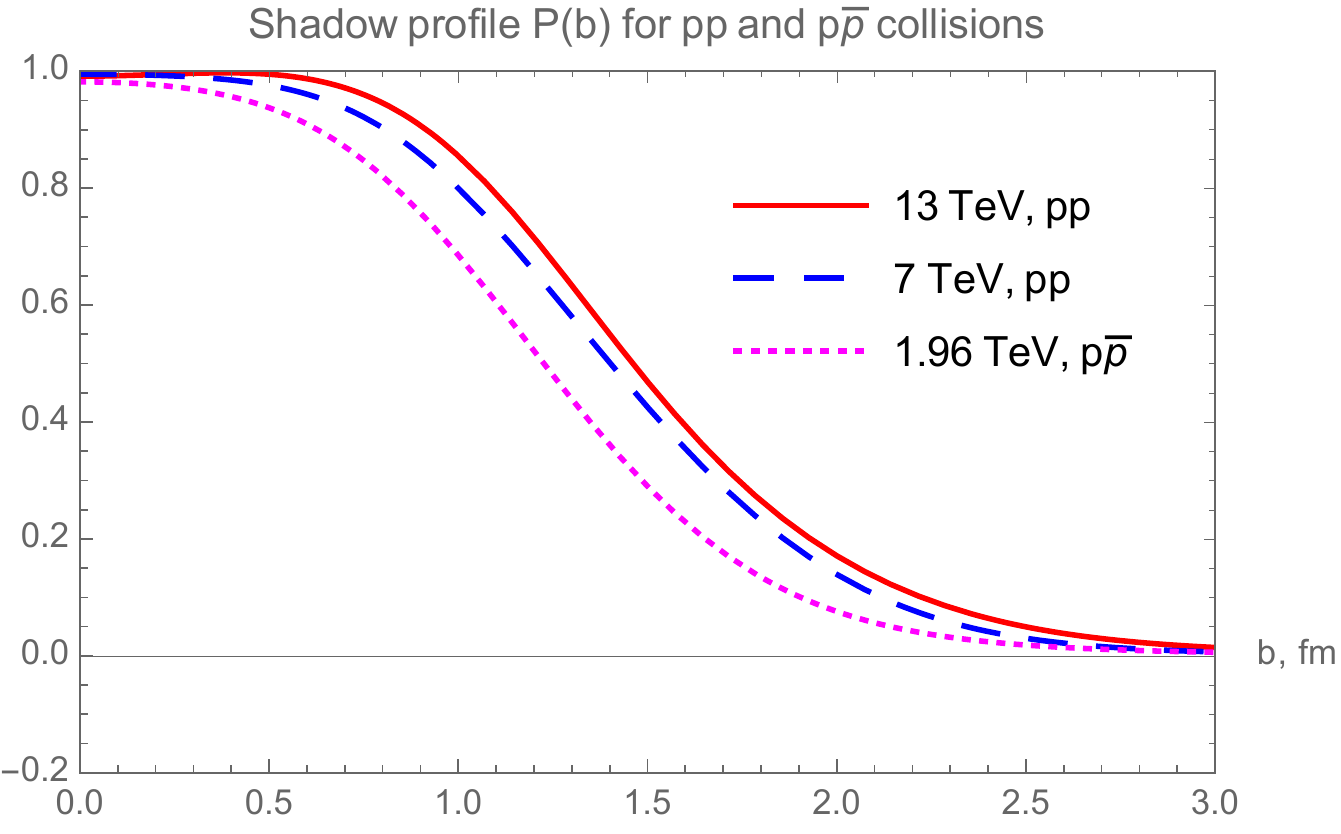}}
\caption{
The shadow profile $P(b)$  for three distinct energies of $pp$ and $p\bar p$ collisions.
}
\label{f:P}
\end{figure}
Fig.~\ref{f:P} indicates the shadow profile of protons at  $\sqrt{s}=$ 7 and 13 TeV in $pp$ and at $\sqrt{s}=$ 1.96 TeV in $p\bar p$ collisions. A nearly black region opens up in the TeV energy region, where $P(b)\approx 1$. This region increases with increasing colliding energies, and it is surrounded by a  gray ``hair" or ``skin" region, where $P(b)$ drops from its maximum to zero with nearly the same decrease at all energies. 
At very large colliding energies, the protons do become blacker and larger, but they do not become edgier, confirming ref.~\cite{Nemes:2015iia}. 

\section{Outlook}
Interesting further details about the theory of imaging, as applied to the internal structure of the protons at LHC energies by elastic scattering, were summarized and highlighted recently in refs.~\cite{Csorgo:2018ruk,Goncalves:2018nsp,Dremin:2019pza}.



\end{document}